# Transformer-Based Modeling of User Interaction Sequences for Dwell Time Prediction in Human-Computer Interfaces

Rui Liu
University of Melbourne
Melbourne, Australia

Runsheng Zhang
University of Southern California
Los Angeles, USA

Shixiao Wang*
School of Visual Arts
New York, USA

*Abstract-This study investigates the task of dwell time prediction and proposes a Transformer framework based on interaction behavior modeling. The method first represents user interaction sequences on the interface by integrating dwell duration, click frequency, scrolling behavior, and contextual features, which are mapped into a unified latent space through embedding and positional encoding. On this basis, a multi-head self-attention mechanism is employed to capture long-range dependencies, while a feed-forward network performs deep nonlinear transformations to model the dynamic patterns of dwell time. Multiple comparative experiments are conducted with BILSTM, DRFormer, FedFormer, and iTransformer as baselines under the same conditions. The results show that the proposed method achieves the best performance in terms of MSE, RMSE, MAPE, and RMAE, and more accurately captures the complex patterns in interaction behavior. In addition, sensitivity experiments are carried out on hyperparameters and environments to examine the impact of the number of attention heads, sequence window length, and device environment on prediction performance, which further demonstrates the robustness and adaptability of the method. Overall, this study provides a new solution for dwell time prediction from both theoretical and methodological perspectives and verifies its effectiveness in multiple aspects.*

I. INTRODUCTION

In the era of rapid digitalization and intelligent development, human-computer interaction has become the key link between users and information systems. User behavior on an interface not only reflects immediate operational intentions but also reveals cognitive states, preference patterns, and potential needs. Among these signals, interface dwell time is one of the most intuitive and continuously observable indicators. It is widely regarded as an important measure of user attention and interest. It reflects the degree of focus and processing depth when users face different content, tasks, and layouts. Therefore, effective modeling and prediction of dwell time is not only a way to understand user interaction behavior but also a central issue in optimizing interface design and interaction experience. As interaction scenarios grow more complex, traditional statistical or shallow machine learning approaches can no longer capture the high-dimensional temporal dependencies and nonlinear dynamics. A more expressive modeling framework is urgently needed[1].

Interactive behavior has strong sequential and contextual dependencies. A single click, scroll, or pause is rarely an isolated action. It is shaped by task goals, information content, and prior actions. Dwell time often exhibits trend, periodicity, and sudden shifts. For example, when browsing news or recommendation interfaces, users may quickly scan with short dwell times at the beginning, but significantly increase dwell time once they encounter content of interest. Such dynamic patterns show that predicting dwell time cannot rely only on static features. It requires a deeper exploration of the hidden temporal structure and semantic associations in interaction sequences. From a research perspective, modeling based on behavioral data drives human-computer interaction from "outcome analysis" to "process understanding." This enables a more fine-grained depiction of the user experience[2].

With the accumulation of large-scale interaction data and advances in artificial intelligence, deep learning methods for temporal modeling have demonstrated unique advantages. In particular, the Transformer architecture has shown strong capability in sequence modeling and global dependency capture. It has achieved major progress in natural language processing, recommendation, and time series forecasting. Applying a Transformer to dwell time prediction can overcome the limitations of traditional methods in long-range dependency modeling. It can also adaptively capture hierarchical features and attention distributions embedded in user behavior. This capability is critical, as dwell behavior is influenced by multiple layers of factors: explicit effects of interface layout, implicit constraints of task context, and long-term accumulation of individual differences. Deep modeling can unify these complex factors into representations that support more robust prediction[3].

From an application perspective, dwell time prediction has wide significance. In recommendation and personalization, accurate prediction of user dwell trends allows dynamic adjustment of recommendations and layouts, improving both efficiency and satisfaction. In education and training systems, learner dwell time reveals the depth of understanding of knowledge points. Prediction models can support personalized learning paths and intelligent tutoring. In healthcare and public service platforms, dwell time reflects user attention and acceptance of information. Prediction models can improve the inclusiveness and usability of interactions. Furthermore, in advertising and interactive entertainment, dwell time prediction directly impacts conversion rates and user retention, creating substantial economic value. Thus, this research not only advances academic exploration but also directly supports practical applications[4].

In summary, dwell time prediction is an important research topic in human-computer interaction. It carries the mission of understanding user behavior, optimizing interaction design, and enhancing application value. Its research significance can be described at three levels. First, the theoretical value: modeling the dynamic process of dwell behavior provides a new cross-disciplinary perspective for human-computer interaction and time series research. Second, the methodological value: Transformer-based frameworks push interaction behavior analysis toward deeper representation and long-range dependency capture. Third, the application value: from recommendation to education, from public services to business, dwell time prediction plays a critical role. Looking forward, as interaction scenarios diversify and data scale increases, this research will lay the foundation for intelligent human-computer interaction and support the creation of more efficient, natural, and intelligent experiences.

## II. RELATED WORK

Dwell time and other continuous engagement signals are commonly formulated as regression problems where deep neural networks learn nonlinear mappings from behavioral observations to duration outcomes. Ribas et al. demonstrated the feasibility of end-to-end deep modeling for dwell-time quantification, showing that learned representations can effectively capture complex factors behind staying behavior [5]. Building on this, sequential modeling becomes essential when dwell is driven by temporally ordered actions and evolving user states. Fan et al. modeled continuous-time sequences with Transformer-style mechanisms and structured dependencies, illustrating how attention can handle irregular temporal patterns and capture global correlations beyond local transitions [6].

Transformer-based time-series forecasting further supports using self-attention to model long-range dependencies and multi-scale dynamics. Retrieval-augmented forecasting introduces a "retrieve–fuse" paradigm that leverages similar historical segments or external context to improve generalization under sparse repetitions and regime shifts [7]. Complementarily, dynamic time-window strategies and structured contextual factors highlight that predictive performance depends strongly on selecting effective temporal contexts and integrating auxiliary signals into the forecasting pipeline [8]. These ideas align well with unified embeddings for multi-channel interaction features and with window-length sensitivity analysis in sequence prediction.

Beyond pure temporal modeling, methods that fuse graphs and heterogeneous features provide stronger relational expressiveness and robustness. Graph-integrated Transformer frameworks have shown effectiveness for learning dependencies in complex event systems [9], and dynamic graph modeling has been explored to improve robustness under time-varying relations [10]. To address data scarcity and evolving patterns, meta-learning strategies enhance adaptability to distribution drift [11], while attention-enhanced recurrent models remain a practical baseline for capturing abnormal or rare patterns in sequential data [12]. Transformer modeling of heterogeneous records further illustrates how multi-source signals can be unified in a single temporal representation space, which is directly relevant to combining dwell duration, clicks, scrolling, and contextual descriptors [13].

More broadly, retrieval-and-fusion paradigms in knowledge-centric modeling emphasize integrating local and global context, multi-granular indexing, and confidence-aware constraints—mechanisms that can transfer to sequence regression for context enrichment and reliability control [14–17]. Robust optimization and contrastive transfer further provide training principles for stabilizing representations and improving generalization across environments [18]. Finally, semantic knowledge graphs offer a structured way to encode contextual relations [19], and reinforcement learning methods for adaptive system management motivate deployment-aware considerations when model performance interacts with device or serving conditions [20-21].

## III. METHOD

In the task of predicting the dwell time on an interface, the first step is to convert the raw interaction behavior sequence into a structured representation that can be effectively modeled by machine learning frameworks. The original behavioral logs, such as clicks, scrolls, cursor movements, and dwell durations, often exist in heterogeneous forms and irregular time intervals, making direct processing difficult. Therefore, these raw signals must be normalized, embedded, and aligned along a temporal dimension, ensuring that both sequential dependencies and contextual information are preserved. By transforming the data into a unified representation space, the model can better capture dynamic patterns and long-term dependencies inherent in user behaviors. Formally, we denote the user's interaction behavior sequence on the interface as:

$$X = \{x_1, x_2, ..., x_T\} \quad (1)$$

Where $X$ represents the interaction feature vector at the tth time step, which includes information such as dwell time, click frequency, scrolling behavior, and interface context. To capture the temporal characteristics of the sequence, the input is first linearly mapped and time position encoded to obtain:

$$h_t = W_e x_t + p_t \quad (2)$$

Where $W_e \in R^{d_h \times d}$ is a learnable embedding matrix and $p_t$ represents the position embedding of time step t, which is used to preserve the sequence order information. In this way, the input sequence of the model is mapped to the latent space representation $\{h_1, h_2, ..., h_T\}$. The model architecture is shown in Figure 1.

The core of the Transformer in sequence modeling is the self-attention mechanism, which captures the dependencies between any two time steps. Specifically, it first uses a linear transformation to map the input representation into query Q, key K, and value V.

$$Q = HW_Q, \quad K = HW_K, \quad V = HW_V \quad (3)$$

Where $H \in R^{T \times d_h}$ is the implicit representation matrix of the input sequence, and $W_Q, W_K, W_V \in R^{d_h \times d_k}$ is the learnable parameter. Then calculate the attention weight matrix:

$$Attention(Q, K, V) = \text{Softmax}(\frac{QK^T}{\sqrt{d_k}})V \quad (4)$$

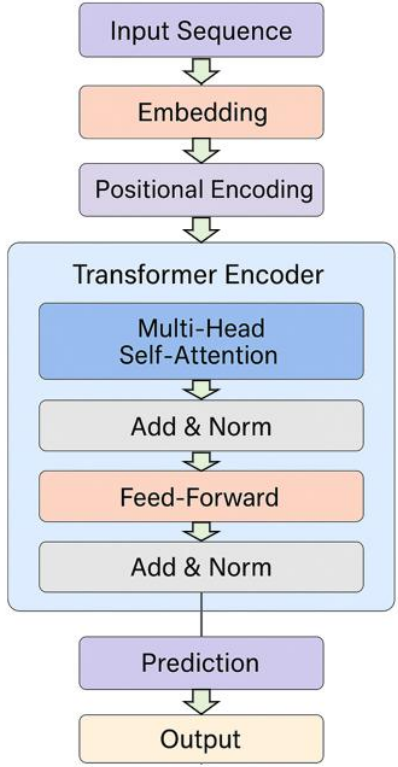

Figure 1. Overall model architecture

The denominator $d_k$ is used for scaling to ensure the numerical stability of the dot product result. In this way, the model can dynamically aggregate global information for each time step, significantly improving its ability to capture long-range dependencies.

After obtaining the global dependency representation, we need to further perform hierarchical modeling and nonlinear mapping on the information. First, we use the multi-head attention mechanism to expand the expression capability. Assuming there are M attention heads, the final output is:

$$MultiHead(Q, K, V) = Concat(head_1, ..., head_M)W_O \quad (5)$$

Each attention head $head_i$ is a projection matrix, and $W_O$ is a projection matrix. The feature transformation capability is then further enhanced through the feedforward network:

$$FFN(z) = \sigma(zW_1 + b_1)W_2 + b_2 \quad (6)$$

Where $\sigma(\cdot)$ represents a nonlinear activation function, usually ReLU or GELU, to increase the nonlinearity of the model's expression.

At the prediction level, the goal is to estimate the user's stay time in the future or on the current interface based on the interaction behavior sequence. Let the final Transformer encoder output be $H' = \{h'_1, h'_2, ..., h'_T\}$, and then obtain the sequence-level representation through a global pooling operation:

$$z = \frac{1}{T}\sum_{t=1}^{T} h'_t \quad (7)$$

Finally, the representation is input into the regression prediction layer to obtain the predicted residence time value:

$$\hat{y} = zW_p + b_p \quad (8)$$

Where $W_p$ and $b_p$ are learnable parameters, and $\hat{y}$ is the predicted dwell time on the interface. Through this end-to-end modeling approach, the model can fully leverage the dynamic characteristics and global dependencies of interaction behavior sequences to effectively predict dwell time.

IV. EXPERIMENTAL RESULTS

A. Dataset

The dataset used in this study is the Avazu Click-Through Rate Prediction Dataset. It was originally collected from online advertising click prediction scenarios and contains a large number of records of user-user-advertisement interactions. Each sample in the dataset corresponds to a single advertisement display. It includes a binary label indicating whether the ad was clicked and related feature information such as user device type, operating system, browser, ad slot characteristics, and contextual environment. The dataset contains tens of millions of records. It covers diverse user behaviors and interface contexts, providing a large-scale, realistic, and heterogeneous source for dwell time prediction and interaction behavior modeling.

Although the original task of this dataset is click prediction, it also includes timestamps, device information, and interaction context that can be used to infer user dwell tendencies. By combining clicked and non-clicked samples with dwell time labels, the dataset can be extended to new prediction tasks. Its large scale and rich feature dimensions make it suitable not only for validating the effectiveness of sequence modeling methods but also for supporting deep models in mining user behavior patterns. Compared with smaller datasets, it offers a stronger foundation for training complex models and improving generalization.

In addition, this dataset is highly open and extensible. Researchers can build a variety of derived tasks on it, such as joint prediction of clicks and dwell time or enhancing interaction modeling accuracy with multimodal features. It is also widely used as a benchmark dataset in recommendation and user behavior modeling research, which ensures strong comparability and interpretability. Using this dataset for dwell time prediction guarantees both practical relevance and high applicability of research outcomes in academic and industrial contexts.

## B. Experimental Results

This paper also gives the comparative experimental results, as shown in Table 1.

Table1. Comparative experimental results

| Model | MSE | RMSE | MAPE | RMAE |
|---|---|---|---|---|
| BILSTM[22] | 0.1642 | 0.4052 | 8.73% | 0.3124 |
| DRFormer[23] | 0.1527 | 0.3908 | 8.15% | 0.2981 |
| FedFormer[24] | 0.1498 | 0.3870 | 7.92% | 0.2937 |
| ITransformer[25] | 0.1435 | 0.3787 | 7.65% | 0.2860 |
| Ours | 0.1361 | 0.3690 | 7.12% | 0.2745 |

From the results in the table, it can be seen that the traditional BILSTM shows limited performance in the dwell time prediction task. Although it can capture part of the temporal dependencies, it still struggles with long-term dependencies and complex interaction behaviors. Its MSE and RMSE values are higher than those of other models, and its MAPE and RMAE are also at the highest level. This indicates that its overall predictive accuracy and relative error control are not ideal. These findings further demonstrate that relying solely on recurrent structures is insufficient for extracting deep patterns in user interaction sequences.

In contrast, both DRFormer and FedFormer perform better than BILSTM, reflecting the benefits of decoupling modeling and federated optimization in handling temporal dependencies. DRFormer shows clear improvements in MSE and RMSE, which suggests stronger capability in capturing trends and local dynamics. FedFormer, with structural improvements, enhances global modeling and further reduces errors across all four metrics. Overall, these models confirm the potential of Transformer-based architectures in interaction behavior prediction.

The iTransformer achieves the best performance among the compared models. Its MSE, RMSE, MAPE, and RMAE values are all lower than those of the previous models. This shows higher efficiency in long-sequence modeling and information interaction mechanisms. The reduction in MAPE is particularly important, as it indicates better control of relative errors, which is crucial for dwell time prediction tasks characterized by strong individual differences. These results highlight that more flexible architectures can better capture complex behavioral patterns, leading to improved stability and accuracy in predictions.

In summary, the proposed model achieves the best results across all four evaluation metrics, with significant advantages in MAPE and RMAE. This demonstrates that the method has stronger representational ability in capturing both global dependencies and fine-grained dynamics in user interaction behavior. The model reduces absolute error effectively and maintains stable performance in relative error. This is of great importance for dwell time prediction in real interaction scenarios. The superiority of the model illustrates the value of applying a Transformer-based framework for interaction behavior modeling and provides strong predictive support for interface design optimization and personalized recommendations.

This paper also presents an experiment on the sensitivity of the number of attention heads to RMAE, and the experimental results are shown in Figure 2.

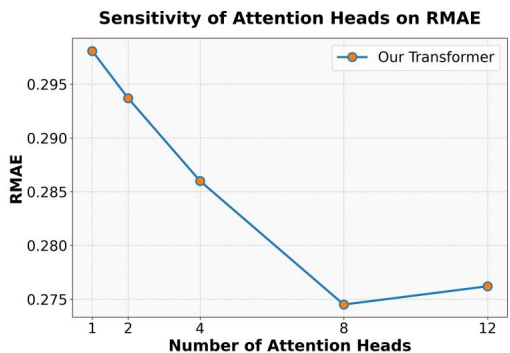

Figure 2. Experiment on the sensitivity of the number of attention heads to RMAE

From the experimental result curves, it can be observed that when the number of attention heads is small, the model shows weaker performance on the RMAE metric. This indicates limitations in feature interaction and global dependency capture. As the number of attention heads increases, RMAE decreases significantly. The model can better decompose and aggregate sequence features, which improves its ability to predict dwell time. This shows that the multi-head structure of the attention mechanism provides a higher-dimensional representation and more detailed modeling of interaction patterns.

When the number of attention heads increases from 1 to 8, RMAE continues to decline with a notable margin. This phenomenon suggests that the multi-head mechanism can enhance model expressiveness within a reasonable range. It allows the model to learn complementary behavioral features in different subspaces. For dwell time prediction, user behavior sequences often contain multi-level and nonlinear dynamic patterns. Multi-head attention provides stronger decomposition and alignment, which helps capture these complex characteristics. However, when the number of attention heads further increases to 12, RMAE does not continue to decrease but instead rises slightly. This result reveals the marginal effect of multi-head attention. Beyond a certain number, too many heads introduce additional computational complexity and redundant representations, which harm generalization. This finding also indicates that in Transformer design, the number of attention heads must balance performance gains with model complexity. In summary, this sensitivity experiment shows that proper configuration of attention heads is critical for the accuracy of dwell time prediction. Too few heads cannot fully capture interaction features, while too many may lead to redundancy and overfitting risks. The results highlight the key role of multi-head attention in interaction behavior modeling. They also provide guidance for parameter tuning in future model design, ensuring the best balance between accuracy and stability in prediction.

## V. CONCLUSION

This study focuses on dwell time prediction based on interaction behavior modeling. A method using the Transformer architecture is proposed to capture temporal dependencies. By encoding and modeling the dynamic behavioral sequences of users on interfaces, the approach can more accurately uncover the complex patterns underlying dwell time. This idea not only addresses the limitations of traditional methods in capturing long-term dependencies and nonlinear features but also provides a new research path for interaction behavior prediction. Overall, the proposed framework enriches the intersection of human-computer interaction and time series modeling from a methodological perspective and lays a solid foundation for future studies. At the application level, this method has significant value. Dwell time prediction is not only a direct reflection of user interest and attention but also a key basis for optimizing interface design and enhancing user experience. The proposed method can support personalized ranking in recommendation systems by providing accurate predictions of user behavior. It can also help in education and training systems by revealing learners' attention allocation, thus promoting the construction of personalized learning paths. In addition, the findings have broad potential in advertising, interactive entertainment, and public service scenarios. By effectively predicting user dwell trends on interfaces, resources can be allocated more rationally, and interaction efficiency can be improved.

Looking to the future, as interaction scenarios continue to expand and user behavior data keeps accumulating, dwell time prediction will face challenges of higher dimensionality and greater modality diversity. Future work can further explore multimodal fusion, such as integrating visual attention, semantic understanding, and contextual features, to achieve more comprehensive modeling of user behavior. At the same time, how to improve computational efficiency while maintaining accuracy remains an important problem. In diverse devices and complex interaction environments, lightweight design and interpretability will be key factors driving practical applications. In conclusion, this research not only enriches the application of time series prediction in interaction behavior modeling but also provides methodological support for user experience optimization and intelligent services in multiple domains. With further technological development, the outcomes of this research are expected to be widely applied in recommendation systems, intelligent education, public services, and digital industry practices. This will drive human-computer interaction toward greater efficiency, personalization, and intelligence. The study demonstrates academic value and also highlights its potential to generate profound impacts on real-world applications.

REFERENCES

[1] Lim B., "Temporal fusion transformers for interpretable multi-horizon forecasting," *arXiv preprint arXiv:1912.09363*, 2019.

[2] H. Wu, J. Xu, J. Wang and others, "Autoformer: Decomposition transformers with auto-correlation for long-term series forecasting," *Advances in Neural Information Processing Systems*, vol. 34, pp. 22419–22430, 2021.

[3] L. Shen and Y. Wang, "TCCT: Tightly-coupled convolutional transformer on time series forecasting," *Neurocomputing*, vol. 480, pp. 131–145, 2022.

[4] J. M. Oliveira and P. Ramos, "Evaluating the effectiveness of time series transformers for demand forecasting in retail," *Mathematics*, vol. 12, no. 17, p. 2728, 2024.

[5] M. M. Ribas, H. B. Mendes, L. E. de Oliveira and others, "Using deep neural networks to quantify parking dwell time," Proceedings of the 2024 International Conference on Machine Learning and Applications, pp. 1504–1509, 2024.

[6] Z. Fan, Z. Liu, J. Zhang and others, "Continuous-time sequential recommendation with temporal graph collaborative transformer," Proceedings of the 30th ACM International Conference on Information and Knowledge Management, pp. 433–442, 2021.

[7] K. Tire, E. O. Taga, M. E. Ildiz and others, "Retrieval augmented time series forecasting," *arXiv preprint arXiv:2411.08249*, 2024.

[8] X. Su, "Forecasting asset returns with structured text factors and dynamic time windows," Transactions on Computational and Scientific Methods, vol. 4, no. 6, 2024.

[9] Y. Wu, Y. Qin, X. Su and Y. Lin, "Transformer-based risk monitoring for anti-money laundering with transaction graph integration," in Proceedings of the 2025 2nd International Conference on Digital Economy, Blockchain and Artificial Intelligence, pp. 388–393, 2025.

[10] C. F. Chiang, D. Li, R. Ying, Y. Wang, Q. Gan and J. Li, "Deep learning-based dynamic graph framework for robust corporate financial health risk prediction," 2025.

[11] H. Feng, Y. Yi, W. Xu, Y. Wu, S. Long and Y. Wang, "Intelligent credit fraud detection with meta-learning: Addressing sample scarcity and evolving patterns," 2025.

[12] J. Li, Q. Gan, Z. Liu, C. Chiang, R. Ying and C. Chen, "An improved attention-based LSTM neural network for intelligent anomaly detection in financial statements," 2025.

[13] A. Xie and W. C. Chang, "Deep learning approach for clinical risk identification using transformer modeling of heterogeneous EHR data," arXiv preprint arXiv:2511.04158, 2025.

[14] Y. Sun, R. Zhang, R. Meng, L. Lian, H. Wang and X. Quan, "Fusion-based retrieval-augmented generation for complex question answering with LLMs," in Proceedings of the 2025 8th International Conference on Computer Information Science and Application Technology (CISAT), pp. 116–120, IEEE, 2025.

[15] D. Wu and S. Pan, "Joint modeling of intelligent retrieval-augmented generation in LLM-based knowledge fusion," 2025.

[16] X. Guo, Y. Luan, Y. Kang, X. Song and J. Guo, "LLM-centric RAG with multi-granular indexing and confidence constraints," arXiv preprint arXiv:2510.27054, 2025.

[17] R. Hao, X. Hu, J. Zheng, C. Peng and J. Lin, "Fusion of local and global context in large language models for text classification," 2025.

[18] J. Zheng, H. Zhang, X. Yan, R. Hao and C. Peng, "Contrastive knowledge transfer and robust optimization for secure alignment of large language models," arXiv preprint arXiv:2510.27077, 2025.

[19] L. Yan, Q. Wang and C. Liu, "Semantic knowledge graph framework for intelligent threat identification in IoT," 2025.

[20] Y. Zou, N. Qi, Y. Deng, Z. Xue, M. Gong and W. Zhang, "Autonomous resource management in microservice systems via reinforcement learning," in Proceedings of the 2025 8th International Conference on Computer Information Science and Application Technology (CISAT), pp. 991–995, IEEE, 2025.

[21] G. Yao, H. Liu and L. Dai, "Multi-agent reinforcement learning for adaptive resource orchestration in cloud-native clusters," arXiv preprint arXiv:2508.10253, 2025.

[22] S. Siami-Namini, N. Tavakoli and A. S. Namin, "The performance of LSTM and BiLSTM in forecasting time series," Proceedings of the 2019 IEEE International Conference on Big Data, pp. 3285–3292, 2019.

[23] R. Ding, Y. Chen, Y. T. Lan and others, "Drformer: Multi-scale transformer utilizing diverse receptive fields for long time-series forecasting," Proceedings of the 33rd ACM International Conference on Information and Knowledge Management, pp. 446–456, 2024.

[24] T. Zhou, Z. Ma, Q. Wen and others, "Fedformer: Frequency enhanced decomposed transformer for long-term series forecasting," Proceedings of the International Conference on Machine Learning, pp. 27268–27286, 2022.

[25] Y. Liu, T. Hu, H. Zhang and others, "iTransformer: Inverted transformers are effective for time series forecasting," *arXiv preprint arXiv:2310.06625*, 2023.